\documentclass[
superscriptaddress,
 amsmath,amssymb,
 aps, physrev,
 twocolumn,
]{revtex4-2}

\usepackage{graphicx}
\usepackage{dcolumn}
\usepackage{bm}
\usepackage{hyperref}
\usepackage{xcolor}
\usepackage{booktabs}
\usepackage{amsmath}
\usepackage{mathtools}

\begin{document}


\title{\textbf{Decomposition of angular momentum projected nuclear wave function} 
}%
 
\author{Wen Chen}
\author{Zhan-Jiang Lian}
\author{Xue-Wei Li}
\author{Xin-Yang Xia}
\author{Zi-Yang He}
\author{Ke-Zheng Ruan}
\author{Zao-Chun Gao}
\email{zcgao@ciae.ac.cn}
\affiliation{China Institute of Atomic Energy P. O. Box 275(10) Beijing 102413 China}

\date{\today}

\begin{abstract}
\noindent\textbf{Abstract:} Angular momentum projection is a basic technique in constructing nuclear wave functions with good spins. Traditionally, a projected nuclear wave function is expressed in terms of the bases built by performing the angular momentum projection directly on reference states for the whole nuclear system. Alternatively, one can construct nuclear wave function with another kind of projected bases, called as the coupled projected bases, which are generated by first performing the angular momentum projections  on the reference states for neutrons and protons, respectively, then coupling the neutron projected states with the proton ones via Clebsch–Gordon coefficients. In the present work, we derive a new identity, which provides a decomposition of the conventional angular momentum projected nuclear wave function in terms of the coupled projected bases. This decomposition offers direct insight into the underlying structure of nuclear states. To show this point, we present the  decompositions of variation after projection shell model (VAPSM) wave functions for the ground states in some $sd$ shell nuclei. It is interesting to see that even for the ground states in even-even nuclei, the nucleons are not fully paired. Finally, we demonstrate that the VAPSM wave function can be further improved by adopting the coupled projected bases.

\vspace{10pt}


\end{abstract}

\maketitle



\section{Introduction}\label{sec:1}

Nuclei are quantum many-body systems composed of protons and neutrons. Exactly solving such quantum many-body problems remains a serious challenge. To date, a variety of approximate quantum many-body methods have been developed. Among them, the Hartree–Fock (HF) approximation \cite{Hartree,Fock} is one of the most important approaches and has been widely applied to many quantum many-body systems, including nuclei \cite{Ring}. In this method, nucleons in a nucleus are assumed to move independently in a common mean field. Although residual interactions among nucleons are neglected, the HF approximation has been very successful in reproducing various global properties of nuclei.
The Hartree–Fock–Bogoliubov (HFB) approximation generalizes the HF approach by introducing the Bogoliubov transformation \cite{Bogolyubov1,GHF}, and it is particularly powerful in describing pairing correlations in nuclei. However, these mean-field approximations generally break fundamental symmetries, such as rotational invariance and reflection symmetry, which are expected to be conserved in nuclear systems.

A natural way to restore these symmetries is to perform quantum-number projections on reference states, such as the HF or HFB vacuum states \cite{Ring}. Particle-number projection can be applied to HFB vacuum states, while angular-momentum projection (AMP) and parity projection can be applied to both HF and HFB ones. Parity projection is relatively simple. By contrast, angular-momentum projection is computationally more demanding because it involves a three-dimensional integration over Euler angles.

Despite its computational complexity, the angular-momentum projection technique has been widely used in the development of various beyond-mean-field methods \cite{Hara&Sun,Sun16,Zhao16,Yao09,Hagen2022,Yao22,Sunxx21,Bally19,Sheikh21,Gao2009,Lian2023,luxiao2023,Cui22}. Moreover, it plays a crucial role in high-quality approximate shell-model approaches, such as the Monte Carlo shell model (MCSM) \cite{OTSUKA2001,Utsuno2013,MCSM1}, the VAMPIR method \cite{vampir}, the quasi-particle vacuum shell model (QVSM) \cite{Shimizu2022}, the discrete nonorthogonal shell model (DNO-SM) \cite{Dao22}, and the variation-after-projection shell model (VAPSM) \cite{Gao2015,Wangjiaqi2018,GAO2022,Lian24}.

Traditionally, nuclear wave functions can be constructed by performing angular-momentum projection on reference states for the entire nuclear system, as implemented in Refs.~\cite{OTSUKA2001,Utsuno2013,MCSM1,vampir,Shimizu2022,Dao22,Gao2015,GAO2022,Wangjiaqi2018,Lian24}. This treatment implicitly assumes that neutrons and protons move coherently as a whole, with no relative collective motion between the two subsystems.

However, collective modes involving relative motion between neutrons and protons, such as the scissors mode, were already studied in the 1970s \cite{Iudice&Palumbo1978} and were later confirmed experimentally \cite{Bohle1984}. The scissors mode has been successfully reproduced within the projected shell model \cite{Sun1998,Lv2022,chenfq2024}, where new projected bases were introduced by coupling the angular-momentum–projected wave functions of the neutrons and protons via Clebsch–Gordon coefficients. Owing to the additional degree of freedom associated with the scissors mode, it is expected that shell-model approximations can be further improved by adopting such coupled projected bases.

In the present work, we expand the calculated VAPSM wave functions in terms of the coupled projected bases analogous to those employed in the study of the scissors mode \cite{Sun1998,Lv2022}. This decomposition allows us to extract the distributions of neutron and proton angular momenta, thereby providing direct insight into the microscopic structure of nuclear states. Furthermore, we demonstrate that the VAPSM description can be systematically improved by incorporating these coupled basis states.

This paper is organized as follows. Section~\ref{identity} presents a general identity associated with angular-momentum projection operators. Section~\ref{decomposition} describes the decomposition of VAPSM wave functions in terms of neutron and proton angular momenta. Section~\ref{imvap} demonstrates the improvement of the VAPSM achieved with the new coupled projected bases. Finally, a summary and outlook are given in Sec.~\ref{summary}.

\section{A general identity}\label{identity}

Here, we call a state to be projected as a reference state. This reference state can be a Slater determinant (SD), as used in the VAPSM, or other complicated wave function without good angular momentum. One can perform the angular momentum projection on a reference state $|\Phi\rangle$, and get a projected state with good spin,
\begin{eqnarray}\label{pj}
|\Psi_{JM}(K)\rangle=N_{JK}P^J_{MK}|\Phi\rangle,
\end{eqnarray}
where $N_{JK}=\frac{1}{\sqrt{\langle\Phi|P^J_{KK}|\Phi\rangle}}$ is the normalization factor, so that $\langle\Psi_{JM}(K)|\Psi_{JM}(K)\rangle=1$. $P^J_{MK}$ is the angular momentum projection operator,
\begin{eqnarray}
 P^J_{MK}=\frac{2J+1}{8\pi^2}\int d\Omega D^{J*}_{MK}(\Omega)\hat R(\Omega),
\end{eqnarray}
where $D^{J}_{MK}(\Omega)=\langle JM|\hat R(\Omega)|JK\rangle$ and $ 
\hat R(\Omega)$ is the rotational operator $\hat R(\Omega)=e^{-\frac{i\alpha  \hat J_z}{\hbar}}e^{-\frac{i\beta  \hat J_y}{\hbar}}e^{-\frac{i\gamma \hat J_z}{\hbar}}$. $\Omega$ stands for the Euler angles, $\alpha$, $\beta$ and $\gamma$. By definition, one can also find that \cite{Ring},
\begin{eqnarray}\label{pjsum}
 P^J_{MK}=\sum_\eta|\eta JM\rangle\langle\eta JK|,
\end{eqnarray}
where all $|\eta JM\rangle$ states form a complete orthonormalized set in the Hilbert space and $\eta$ represents all quantum numbers except for the spin $J$ and the magnetic quantum number $M$.
This immediately tells us that,
\begin{eqnarray}\label{cplt}
\sum_{JK} P^J_{KK}=\sum_{\eta JK}|\eta JK\rangle\langle\eta JK|=1.
\end{eqnarray}

For convenience, we first assume that the reference state $|\Phi\rangle$ for an entire nuclear system is the product of a proton reference state, $|\Phi^{\pi}\rangle$, and a neutron reference state, $|\Phi^{\nu}\rangle$, namely,
\begin{eqnarray}\label{rf}
|\Phi\rangle=|\Phi^\pi\rangle|\Phi^\nu\rangle.
\end{eqnarray}
Similar to Eq.(\ref{pj}), one can separately perform the angular momentum projections on $|\Phi^{\pi}\rangle$ and on $|\Phi^{\nu}\rangle$, i.e.,
\begin{eqnarray}\label{pjnp}
|\Psi_{J_{\pi}M_{\pi}}(K_{\pi})\rangle&=&N_{J_\pi K_\pi} {P}_{M_{\pi}K_{\pi}}^{J_{\pi}}|\Phi^{\pi}\rangle,\nonumber\\
|\Psi_{J_{\nu}M_{\nu}}(K_{\nu})\rangle&=&N_{J_\nu K_\nu} {P}_{M_{\nu}K_{\nu}}^{J_{\nu}}|\Phi^{\nu}\rangle.
\end{eqnarray}
Then the total nuclear wave functions with good quantum numbers,$J_{\pi},J_{\nu},J,M$, can be constructed as
\begin{eqnarray}\label{cp}
&&|\Psi_{J_{\pi}J_{\nu}JM}(K_\pi,K_\nu)\rangle\\
&=&\sum_{M_\pi M_\nu}\langle J_{\pi}M_{\pi}J_{\nu}M_{\nu}|JM\rangle|\Psi_{J_{\pi}M_{\pi}}(K_{\pi})\rangle
|\Psi_{J_{\nu}M_{\nu}}(K_{\nu})\rangle.\nonumber
\end{eqnarray}

From Eqs.(\ref{cplt}) and (\ref{rf}), one can easily have 
\begin{eqnarray}
&&\sum_{JK}{P}_{KK}^{J}|\Phi\rangle\nonumber\\&=&\left(\sum_{J_{\pi}K_{\pi}}{P}_{K_{\pi}K_{\pi}}^{J_{\pi}}|\Phi^{\pi}\rangle\right)\left(\sum_{J_{\nu}K_{\nu}}{P}_{K_{\nu}K_{\nu}}^{J_{\nu}}|\Phi^{\nu}\rangle\right).
\end{eqnarray}
Notice that ${P}_{K_{\pi}K_{\pi}}^{J_{\pi}}$ can only be applied to $|\Phi^{\pi}\rangle$, and ${P}_{K_{\nu}K_{\nu}}^{J_{\nu}}$ can only be applied to $|\Phi^{\nu}\rangle$. Therefore we have
\begin{eqnarray}
&&\sum_{JK}{P}_{KK}^{J}|\Phi\rangle\nonumber
\\&=&\sum_{J_{\pi}K_{\pi}J_{\nu}K_{\nu}}\frac{|\Psi_{J_{\pi}K_{\pi}}(K_{\pi})\rangle|\Psi_{ J_{\nu}K_{\nu}}(K_{\nu})\rangle}{N_{J_\pi K_\pi}N_{J_\nu K_\nu}}\nonumber\\
&=&\sum_{J_{\pi}K_{\pi}J_{\nu}K_{\nu},J}\frac{\langle J_{\pi}K_{\pi}J_{\nu}K_{\nu}|JK_{\pi}+K_{\nu}\rangle}{N_{J_\pi K_\pi}N_{J_\nu K_\nu}}\\
&&\times|\Psi_{J_{\pi}J_{\nu}JK_{\pi}+K_{\nu}}(K_\pi,K_\nu)\rangle.\nonumber\\
\end{eqnarray}
With the orthogonality among the eigenstates of the angular momentum, one can get
\begin{eqnarray}\label{pjkk}
&&{P}_{KK}^{J}|\Phi\rangle\\&=&\sum_{J_{\pi}J_{\nu}K_{\pi}K_{\nu}}\frac{\langle J_{\pi}K_{\pi}J_{\nu}K_{\nu}|JK\rangle|\Psi_{J_{\pi}J_{\nu}JK}(K_\pi,K_\nu)\rangle}{N_{J_\pi K_\pi}N_{J_\nu K_\nu}}.\nonumber
\end{eqnarray}
By appling Eq.(\ref{pjsum}) to both sides of Eq.(\ref{pjkk}), we have
\begin{eqnarray}
&&{P}_{MK}^{J}|\Phi\rangle\\&=&\sum_{J_{\pi}J_{\nu}K_{\pi}K_{\nu}}\frac{\langle J_{\pi}K_{\pi}J_{\nu}K_{\nu}|JK\rangle|\Psi_{J_{\pi}J_{\nu}JM}(K_\pi,K_\nu)\rangle}{N_{J_\pi K_\pi}N_{J_\nu K_\nu}}.\nonumber
\end{eqnarray}
Using Eq.(\ref{cp}), we then have
\begin{eqnarray}\label{xp}
&&{P}_{MK}^{J}|\Phi\rangle\nonumber\\&=&\sum_{J_{\pi}J_{\nu}K_{\pi}K_{\nu}M_\pi M_\nu}\frac{\langle J_{\pi}K_{\pi}J_{\nu}K_{\nu}|JK\rangle\langle J_{\pi}M_{\pi}J_{\nu}M_{\nu}|JM\rangle}{N_{J_\pi K_\pi}N_{J_\nu K_\nu}}\nonumber\\&&\times|\Psi_{J_{\pi}M_{\pi}}(K_{\pi})\rangle|\Psi_{ J_{\nu}M_{\nu}}(K_{\nu})\rangle
\nonumber\\&=&\sum_{J_{\pi}J_{\nu}K_{\pi}K_{\nu}M_\pi M_\nu}{\langle J_{\pi}K_{\pi}J_{\nu}K_{\nu}|JK\rangle\langle J_{\pi}M_{\pi}J_{\nu}M_{\nu}|JM\rangle}\nonumber\\&&\times{P}_{M_{\pi}K_{\pi}}^{J_{\pi}}{P}_{M_{\nu}K_{\nu}}^{J_{\nu}}|\Phi\rangle.
\end{eqnarray} 
Clearly, the reference state $|\Phi\rangle$ in Eq.(\ref{xp}) can be arbitrary and can be written in a general form, 
\begin{eqnarray}\label{rfg}
|\Phi\rangle=\sum_{ij}f_{ij}|\Phi^\pi_i\rangle|\Phi^\nu_{j}\rangle,
\end{eqnarray}
where $i(j)$ labels the proton (neutron) reference state and the $f_{ij}$ coefficients are arbitrary.
Therefore, we obtain the following new identity
\begin{eqnarray}\label{pjxp0}
&&{P}_{MK}^{J}\nonumber\\&=&\sum_{J_{\pi}J_{\nu}K_{\pi}K_{\nu}M_\pi M_\nu}{\langle J_{\pi}K_{\pi}J_{\nu}K_{\nu}|JK\rangle\langle J_{\pi}M_{\pi}J_{\nu}M_{\nu}|JM\rangle}\nonumber\\&&\times{P}_{M_{\pi}K_{\pi}}^{J_{\pi}}{P}_{M_{\nu}K_{\nu}}^{J_{\nu}}.
\end{eqnarray} 
For convenience, we define the following angular momentum projection operator,
\begin{eqnarray}\label{newpj}
&& P^{J_{\pi}J_{\nu}}_{JM}(K_{\pi}K_{\nu})\nonumber\\
&\equiv&\sum_{M_\pi M_\nu}\langle J_{\pi}M_{\pi}J_{\nu}M_{\nu}|JM\rangle{P}_{M_{\pi}K_{\pi}}^{J_{\pi}}{P}_{M_{\nu}K_{\nu}}^{J_{\nu}},
\end{eqnarray}
and Eq.(\ref{pjxp0}) can be written as
\begin{eqnarray}\label{pjxp}
&&{P}_{MK}^{J}\nonumber\\&=&\sum_{J_{\pi}J_{\nu}K_{\pi}K_{\nu}}\langle J_{\pi}K_{\pi}J_{\nu}K_{\nu}|JK\rangle  P^{J_{\pi}J_{\nu}}_{JM}(K_{\pi}K_{\nu}).
\end{eqnarray} 

\section{Decomposition of the projected wave functions}\label{decomposition}

In general, one can write a projected nuclear wave function as
\begin{eqnarray}\label{pjwf}
|\Psi_{JM}\rangle=\sum_{Ki}f_{Ki} P^J_{MK}|\Phi_i\rangle,
\end{eqnarray}
where the $f_{Ki}$ coefficients satisfy the nomoralization condition so that $\langle\Psi_{JM}|\Psi_{JM}\rangle=1$. Inserting Eq.(\ref{pjxp}) into Eq.(\ref{pjwf}), one can get
\begin{eqnarray}\label{dpj}
|\Psi_{JM}\rangle=\sum_{J_{\pi}J_{\nu}}\sum_{K_{\pi}K_{\nu}i}f_{K i}\langle J_{\pi}K_{\pi}J_{\nu}K_{\nu}|JK\rangle\nonumber\\\times   P^{J_{\pi}J_{\nu}}_{JM}(K_{\pi}K_{\nu})|\Phi_i\rangle.
\end{eqnarray}
Here, we call $P^{J_{\pi}J_{\nu}}_{JM}(K_{\pi}K_{\nu})|\Phi_i\rangle$ as the coupled projected basis. From Eq.(\ref{newpj}), it is seen that $ P^{J_{\pi}J_{\nu}}_{JM}(K_{\pi}K_{\nu})$ is an operator projecting four good quantum numbers, namely, $J_{\pi}$, $J_{\nu}$, $J$ and $M$. This enables us to decompose $|\Psi_{JM}\rangle$ into different ($J_{\pi}$, $J_{\nu}$) components which are orthogonal to one another. The contribution of each ($J_{\pi}$, $J_{\nu}$) component to $|\Psi_{JM}\rangle$ can be evaluated by 
\begin{eqnarray}\label{cbn}
C(J_\pi,J_\nu)=\sum_{K'_{\pi}K'_{\nu}i'K_{\pi}K_{\nu}i}f^*_{K' i'}f_{K i}\langle J_{\pi}K'_{\pi}J_{\nu}K'_{\nu}|JK'\rangle\nonumber\\\times \langle J_{\pi}K_{\pi}J_{\nu}K_{\nu}|JK\rangle\langle\Phi_{i'}|  P^{J_{\pi}}_{K'_{\pi}K_{\pi}} P^{J_{\nu}}_{K'_{\nu}K_{\nu}}|\Phi_i\rangle.
\end{eqnarray}
Clearly, $C(J_\pi,J_\nu)$ should satisfy
\begin{eqnarray}\label{}
\sum_{J_\pi,J_\nu}C(J_\pi,J_\nu)=1.    
\end{eqnarray}

We should mention that similar calculations of $C(J_\pi,J_\nu)$ have been previously made by Otsuka \cite{otsuka1993}. But in his calculation, the intrinsic state is assumed to be axial and has $K=0$ for even-even nuclei. Here, we present Eq.(\ref{cbn})  which is more general and the reference states can be arbitrary. This means Eq.(\ref{cbn}) can be applied to any state in any even-even, odd-mass, or odd-odd nucleus.

In our VAPSM calculation, we have shown that the projected wave function in Eq.(\ref{pjwf}) can be simplified as
\begin{eqnarray}\label{pjwf1}
|\Psi_{JM}(K)\rangle=\sum_{i}f_{i} P^J_{MK}|\Phi_i\rangle,
\end{eqnarray}
without losing good shell model approximation \cite{GAO2022}. Here, $|\Phi_i\rangle$ is a product of a proton SD, $|\Phi^\pi_i\rangle$, and a neutron SD, $|\Phi^\nu_i\rangle$, namely, $|\Phi_i\rangle=|\Phi^\pi_i\rangle|\Phi^\nu_i\rangle$. Accordingly, the $C(J_\pi,J_\nu)$ can be written as
\begin{eqnarray}\label{cbn1}
&&C(J_\pi,J_\nu)\nonumber\\&=&\sum_{K'_{\pi}i'K_{\pi}i}\langle J_{\pi}K'_{\pi}J_{\nu}K-K'_{\pi}|JK\rangle\langle J_{\pi}K_{\pi}J_{\nu}K-K_{\pi}|JK\rangle\nonumber\\&&\times f^*_{i'}f_{i}\langle\Phi^\pi_{i'}|  P^{J_{\pi}}_{K'_{\pi}K_{\pi}}|\Phi^\pi_i\rangle\langle\Phi^\nu_{i'}|  P^{J_{\nu}}_{K-K'_{\pi},K-K_{\pi}}|\Phi^\nu_i\rangle,
\end{eqnarray}
If we take only one reference state $|\Phi\rangle=|\Phi^\pi\rangle|\Phi^\nu\rangle$ then Eq.(\ref{pjwf1}) reduces to Eq.(\ref{pj}).
In this case, the corresponding $C(J_\pi,J_\nu)$ can be calculated by 
\begin{eqnarray}\label{cbn2}
&&C(J_\pi,J_\nu)\nonumber\\
&=&\sum_{K'_{\pi}K_{\pi}}\langle J_{\pi}K'_{\pi}J_{\nu}K-K'_{\pi}|JK\rangle\langle J_{\pi}K_{\pi}J_{\nu}K-K_{\pi}|JK\rangle\nonumber\\&&\times\frac{\langle\Phi^\pi|  P^{J_{\pi}}_{K'_{\pi}K_{\pi}}|\Phi^\pi\rangle\langle\Phi^\nu| P^{J_{\nu}}_{K-K'_{\pi},K-K_{\pi}}|\Phi^\nu\rangle}{\langle\Phi| P^J_{KK}|\Phi\rangle}.
\end{eqnarray}

 For even-even nuclei, their ground states always have $J=0$ spin  and positive parity.
The corresponding $C(J_\pi,J_\nu)$ can be written explicitly
\begin{eqnarray}\label{}
    C(J_\pi,J_\nu)&=&\frac{1}{\langle\Phi|{P}_{00}^{0}|\Phi\rangle}\sum_{KK'}\frac{(-1)^{K'-K}\delta_{J_\pi J_\nu}}{2J_\pi+1}\nonumber\\
&&\langle\Phi^{\pi}|{P}_{K'K}^{J_\pi}|\Phi^{\pi}\rangle\langle\Phi^{\nu}|{P}_{-K'-K}^{J_\nu}|\Phi^{\nu}\rangle.
\end{eqnarray}

\begin{figure}
    \centering
    \includegraphics[width=1\linewidth]{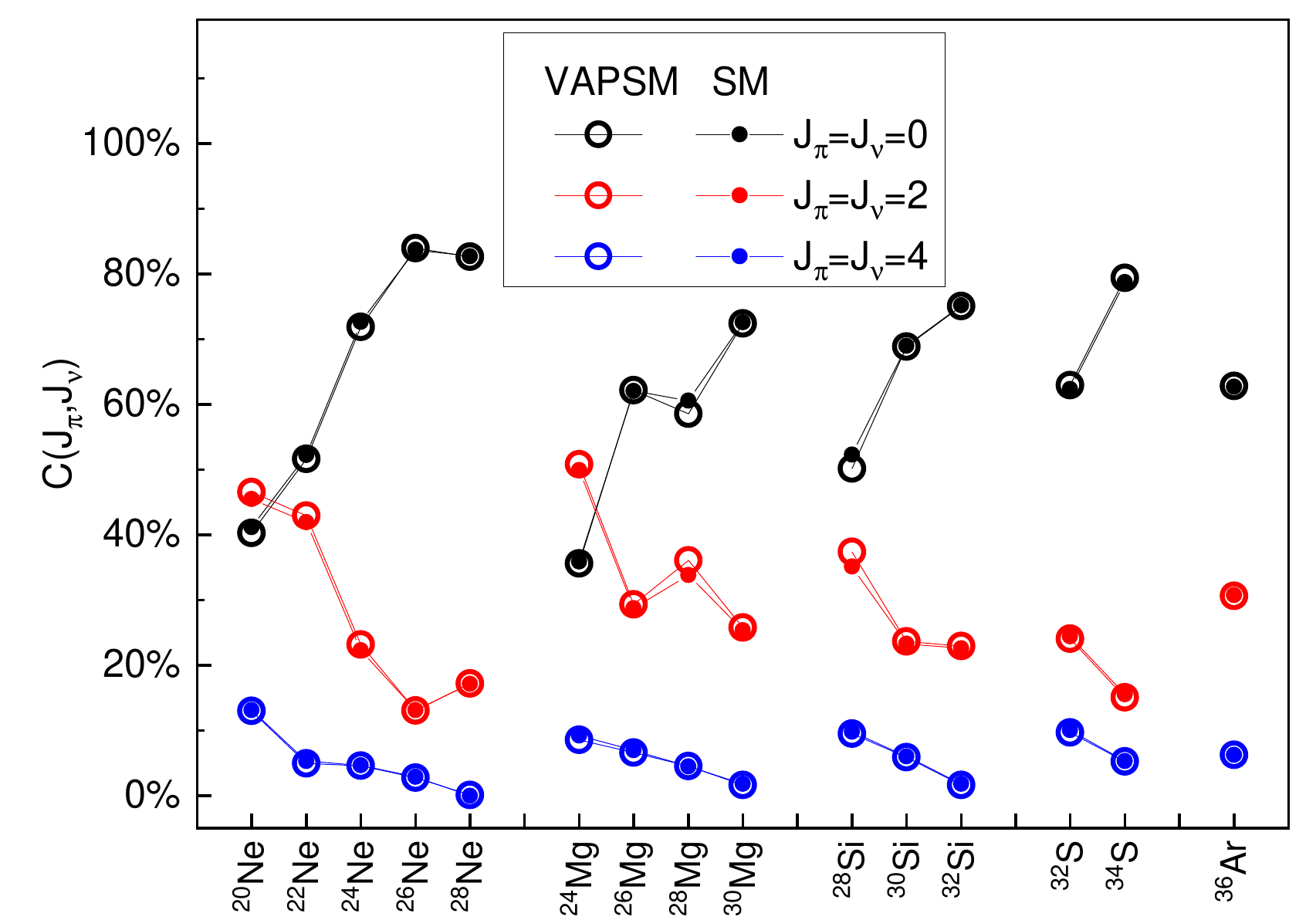}
    \caption{Calculated $C(J_\pi,J_\nu)$ values for the ground states in even-even $sd$ shell nuclei by the shell model and by the VAPSM. The USDB interaction is adopted.}
    \label{fig1}
\end{figure}

Traditionally, when an even-even nucleus is at the ground state,  it is believed that all the nucleons are paired. One may imagine that $C(J_\pi=0,J_\nu=0)$ is expected to be 1 and the other $C(J_\pi,J_\nu)$ values vanish. Unfortunately, shell model calculations have shown that this is not true. To show this point, we calculate the ground states in some even-even $sd$ shell nuclei with both the full shell model and the VAPSM. The adopted shell model Hamiltonian is USDB \cite{usdb06}. The full shell model calculations are performed using the Nushellx code \cite{nushellx}. This shell model code is also capable of calculating the  $C(J_\pi,J_\nu)$ values for any eigenstate of a given Hamiltonian, as shown in Fig.\ref{fig1}.  The VAPSM trial wave function can be taken as Eq.(\ref{pj}). After variation, we calculate the $C(J_\pi,J_\nu)$ values for the fully optimized VAPSM wave functions which are also shown in Fig.\ref{fig1}. 

One can easily see that both SM and VAPSM give very similar $C(J_\pi,J_\nu)$ values. This implies that the approximated VAPSM wave functions are quite close to the corresponding exact SM ones. This also confirms the correctness of the present formulae. However, one can easily see that all the calculated  $C(J_\pi=0,J_\nu=0)$ values are less than 1, while the $C(J_\pi=2,J_\nu=2)$ values are remarkable. For the cases of $^{20}$Ne and $^{24}$Mg, the $C(J_\pi=2,J_\nu=2)$ values are even larger than the $C(J_\pi=0,J_\nu=0)$ values. This indicates that even at the ground states in even-even nuclei, the like nucleons are not fully paired because in the components with $J_\pi=J_\nu=2$, at least two like nucleons are not coupled to spin zero.

Such appearance of the $J_\pi=J_\nu\neq 0$ components is clearly due to the existence of the neutron-proton interaction in the adopted Hamiltonian. This can be understood in the following way. Let's remove the neutron-proton interaction from the adopted Hamiltonian. Then the Hamiltonian can be written as 
\begin{eqnarray}\label{Hnonp}
     H= H^\pi+ H^\nu,
\end{eqnarray}
This means the nuclear system can be separated into two independent subsystems, i.e., the neutron subsystem and the proton subsystem. For each subsystem, the ground state is formed with all like nucleons paired and the corresponding wave function should have zero spin and positive parity. Then the ground state of the total nuclear system is simply the product of the ground states of these two subsystems. This means the ground state for the Hamiltonian of Eq.(\ref{Hnonp}) includes only the  $J_\pi=J_\nu=0$ component. Therefore, it is certain that the neutron-proton interaction leads to the mixture of the $J_\pi=J_\nu\neq 0$ components in the ground states of even-even nuclei.

One can imagine that with the strength of the neutron-proton interaction increasing from zero, the mixing of the  $J_\pi=J_\nu\neq 0$ component might increase from zero accordingly. The $N=Z$ nuclei seem to have the strongest neutron-proton interactions. Indeed, this can be seen in Fig.\ref{fig1} that for the ground states  in $^{20}$Ne, $^{24}$Mg, $^{28}$Si, $^{32}$S and $^{36}$Ar, the weights of the $J_\pi=J_\nu=2$ components are relatively larger. Such $J_\pi=J_\nu=2$ components are even predominant in $^{20}$Ne and $^{24}$Mg. One can also see from Fig.\ref{fig1} that, when the neutron number increases from $N=Z$, $C(J_\pi=0,J_\nu=0)$ roughly increases while $C(J_\pi=2,J_\nu=2)$ roughly decreases. This implies that 
the neutron-proton interaction might become weak with $N$ increasing from $N=Z$.

\begin{figure}
    \centering
    \includegraphics[width=1\linewidth]{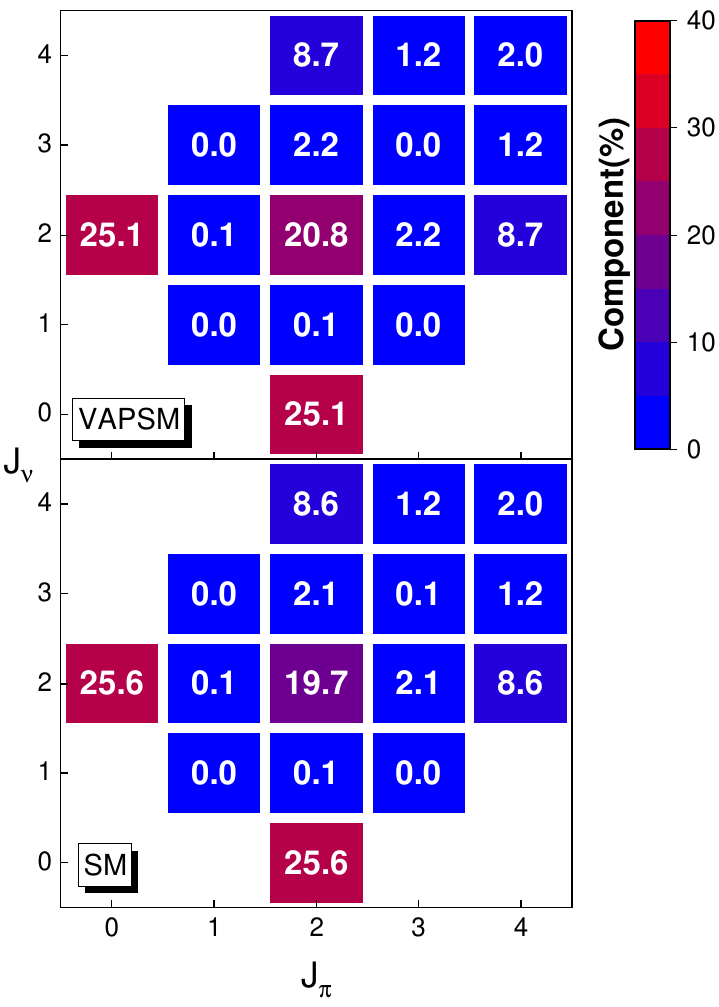}
    \caption{Distribution of the calculated  $C(J_\pi,J_\nu)$ values for the yrast $2^+$ state in $^{24}$Mg by the VAPSM (upper panel) and by the shell model (lower panel). The USDB interaction is adopted.}
    \label{mg24}
\end{figure}
\begin{figure}
    \centering
    \includegraphics[width=1\linewidth]{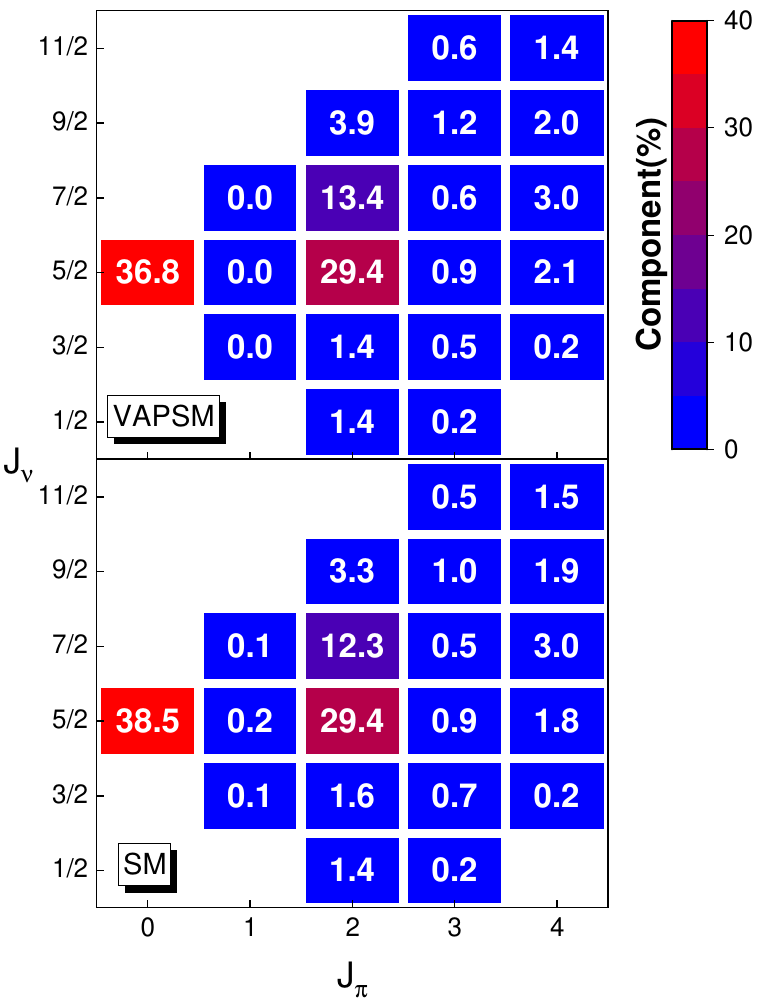}
    \caption{The same as Fig.\ref{mg24} but for the ground $5/2^+$ state in $^{25}$Mg.}
    \label{mg25}
\end{figure}
\begin{figure}
    \centering
    \includegraphics[width=1\linewidth]{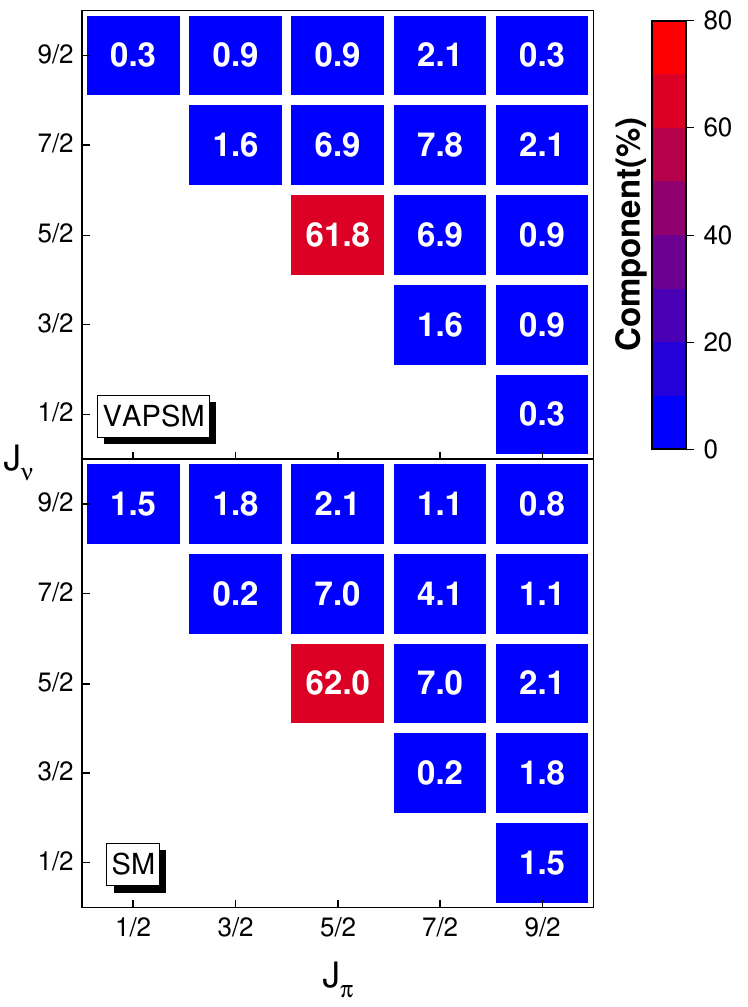}
    \caption{The same as Fig.\ref{mg24} but for the ground $5^+$ state  in $^{26}$Al.}
    \label{al26}
\end{figure}
\begin{figure}
    \centering
    \includegraphics[width=1\linewidth]{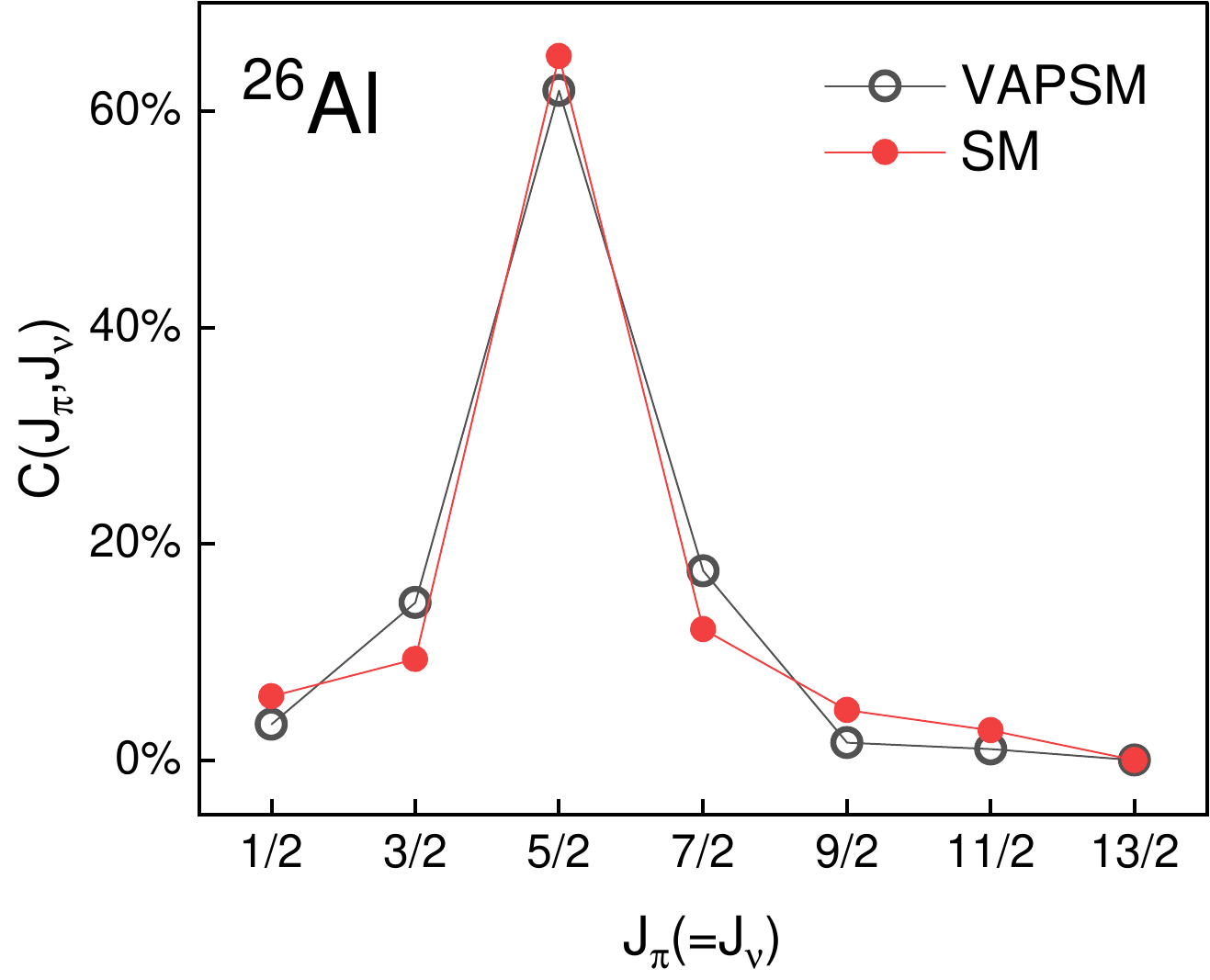}
    \caption{Distribution of the calculated $C(J_\pi,J_\nu)$ values for the yrast $0^+$ state  in $^{26}$Al by the shell model and the VAPSM. The USDB interaction is adopted.}
    \label{al26_0}
\end{figure}
 
For non-zero spin state, the decomposition of the corresponding wave function is more complicated since $J_\pi$ and $J_\nu$ can be different. We perform the same calculations as those in Fig.\ref{fig1} but only change the spin from $J=0$ to $J=2$. The result for the yrast $2^+$ state in $^{24}$Mg is shown in Fig.\ref{mg24}. Again, the distribution of the $C(J_\pi,J_\nu)$ from the the VAPSM is very similar to that from the exact SM. One can see that the main components for the yrast $2^+$ state in $^{24}$Mg are those with $(J_\pi=0,J_\nu=2)$, $(J_\pi=2,J_\nu=0)$ and $(J_\pi=2,J_\nu=2)$. Similar calculations for the odd-mass nucleus $^{25}$Mg and the odd-odd nucleus $^{26}$Al are also performed. The results for their ground states are shown in Figs.\ref{mg25} and \ref{al26}. In Fig.\ref{mg25}, the ground state in $^{25}$Mg is mainly contributed by the $(J_\pi=0,J_\nu=5/2)$ and $(J_\pi=2,J_\nu=5/2)$ components. Definitely, the latter component is also due to the existence of the neutron-proton interactions in the USDB Hamiltonian. 

One may observe that in Figs.\ref{mg24} and \ref{mg25}, when $J_\pi$ or $J_\nu$ is an odd number, the corresponding $C(J_\pi,J_\nu)$ becomes very small. This is also indicated in Fig.\ref{fig1}, where those small $C(J_\pi,J_\nu)$ values are not shown at all. This phenomenon may be caused by the time reversal symmetry. If a deformed intrinsic state has time reversal symmetry, then one can prove that the possible angular momentum projected states with positive parity and $K=0$ are those with even $J$ \cite{Gao02}. This is why in even-even nuclei, the ground state bands only include those states with even $J$. In the lowest proton (neutron) configuration with even particle number, all protons (neutrons),  governed by the Hamiltonian, likely form pairs. One can easily understand that such a configuration may have time reversal symmetry. Therefore, The projected states with $K_\pi^\pi=0^+$ ($K_\nu^\pi=0^+$) can only be those with $J_\pi$ ($J_\nu$) even numbers. Consequently, the $C(J_\pi,J_\nu)$ value should vanish when $J_\pi (J_\nu)$ is an odd number. Small nonzero value of $C(J_\pi,J_\nu)$ with odd $J_\pi$ ($J_\nu$) number implies that the time reversal symmetry  might be slightly broken.

For the ground state in $^{26}$Al, the dominant component in Fig.\ref{al26} is $(J_\pi=5/2,J_\nu=5/2)$, which indicates that the  spins of the neutrons and protons are parallel. One may imagine there is another possibility that those spins can be antiparallel and form a $0^+$ state. Indeed, we calculate the yrast $0^+
$ state in $^{26}$Al and the results are shown in Fig.\ref{al26_0}. Both the shell model and the VAPSM results have shown that such $0^+$ state is also dominated by the  $(J_\pi=5/2,J_\nu=5/2)$ component.

Based on the above analysis, it seems that such decomposition of the angular momentum projected nuclear wave function is very helpful in studying the structure of the calculated nuclear states. This opens the possibility that such decomposition can be applied to any heavy-deformed nucleus once its angular momentum projected wave function is available. 

However, in a heavy deformed nuclear region, the intrude orbits with opposite parity are usually involved in the model space. In this case, the parity projection is adopted and the VAPSM wave function in Eq. (\ref{pjwf1}) can be written as
\begin{eqnarray}\label{pjwf2}
|\Psi^\pi_{JM}(K)\rangle=\sum_{i}f_{i} P^\pi P^J_{MK}|\Phi_i\rangle,
\end{eqnarray}
where $P^\pi=\frac{1}{2}(1+\pi \hat{P})$ is the parity projection operator. $\pi=\pm1$ and $\hat{P}$ is the parity operator. One can easily recognize that 
\begin{eqnarray}
P^\pi=\sum_{\pi_\pi,\pi_\nu}P^{\pi_\pi}P^{\pi_\nu}
\end{eqnarray}
where $P^{\pi_\pi}$ and $P^{\pi_\nu}$ are the parity projection operators for protons and neutrons, respectively. The parities for protons and neutrons, $\pi_\pi$ and $\pi_\nu$, satisfy $\pi_\pi\pi_\nu=\pi$. Therefore, the $C(J_\pi,J_\nu)$ component is split into two parts, namely, 
\begin{eqnarray}\label{cpp}
C(J_\pi,J_\nu)=C(J^+_\pi,J^+_\nu)+C(J^-_\pi,J^-_\nu)
\end{eqnarray}
for $\pi=+1$ state, or
\begin{eqnarray}
C(J_\pi,J_\nu)=C(J^+_\pi,J^-_\nu)+C(J^-_\pi,J^+_\nu)
\end{eqnarray}
for $\pi=-1$ state. Here, $J^{\pi_\tau}_\tau$ stands for the spin and parity for the protons ($\tau=\pi$) or neutrons ($\tau=\nu$). Similar to Eq.(\ref{cbn1}),
$C(J^{\pi_\pi}_\pi,J^{\pi_\nu}_\nu)$ can be calculated by
\begin{eqnarray}\label{cbn3}
&&C(J^{\pi_\pi}_\pi,J^{\pi_\nu}_\nu)\nonumber\\&=&\sum_{K'_{\pi}i'K_{\pi}i}\langle J_{\pi}K'_{\pi}J_{\nu}K-K'_{\pi}|JK\rangle\langle J_{\pi}K_{\pi}J_{\nu}K-K_{\pi}|JK\rangle\nonumber\\&\times&f^*_{i'}f_{i}\langle\Phi^\pi_{i'}|  P^{\pi_\pi}P^{J_{\pi}}_{K'_{\pi}K_{\pi}}|\Phi^\pi_i\rangle\langle\Phi^\nu_{i'}| P^{\pi_\nu} P^{J_{\nu}}_{K-K'_{\pi},K-K_{\pi}}|\Phi^\nu_i\rangle,\nonumber\\
\end{eqnarray}

\begin{figure}
    \centering
    \includegraphics[width=1\linewidth]{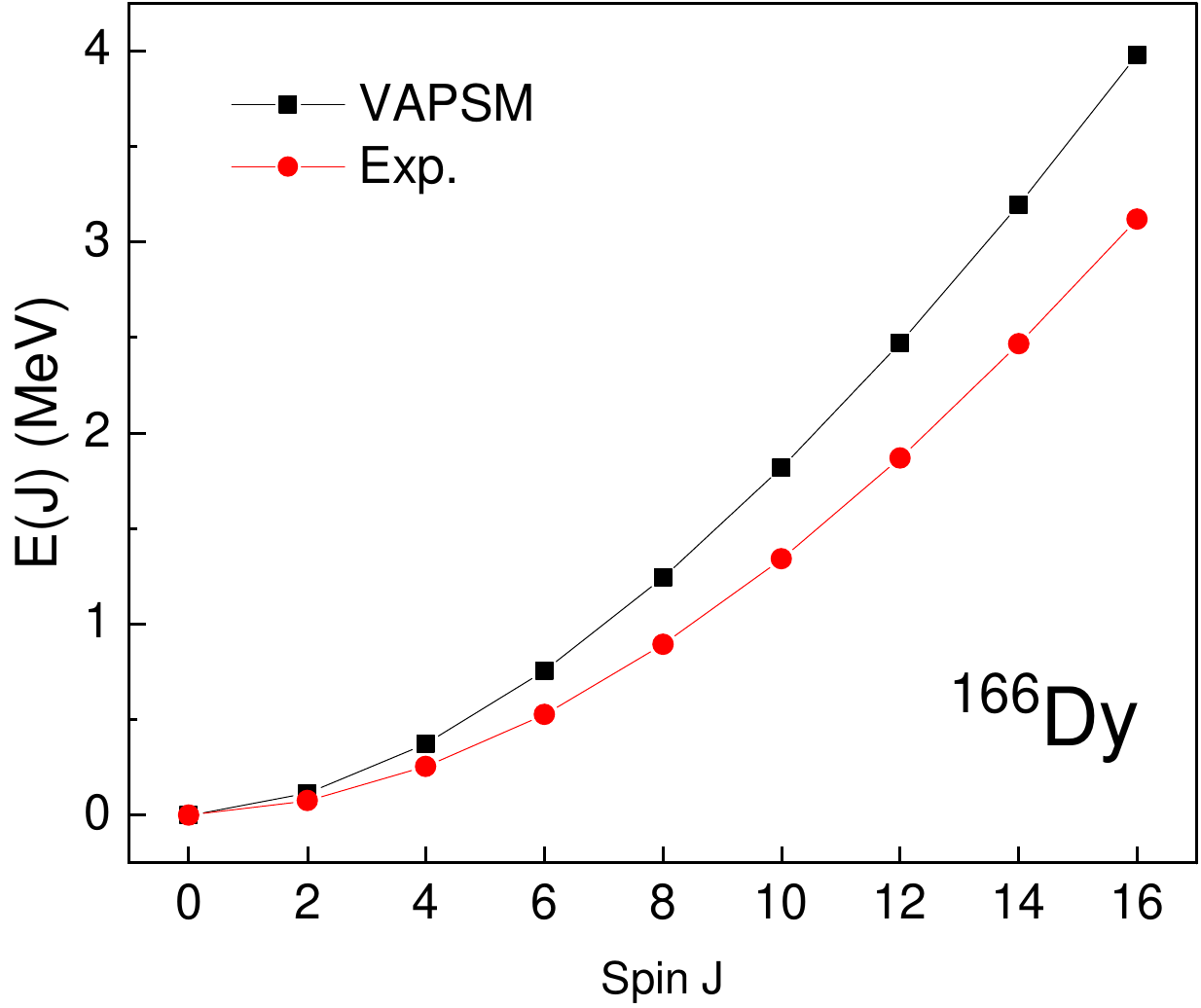}
    \caption{Calculated ground state band energies and the corresponding experimental data in $^{166}$Dy. The jj56pnb interaction is adopted.}
    \label{Dy166_gsb}
\end{figure}

To demonstrate the decomposition in a heavy deformed nucleus, we calculate the ground state band with $K^\pi=0^+$ in $^{166}$Dy. The adopted model space includes all single-particle orbits between the shell closures of $^{132}$Sn and $^{208}$Pb, which is usually called as jj56pn model space. In this model space, the dimension of the configuration space for $^{166}$Dy is as huge as $1.0\times 10^{17}$, which is far outside of the scope of the traditional shell model calculation. The adopted shell model Hamiltonian is jj56pnb \cite{jj56pnb}, which is built for nuclei close to $^{132}$Sn. Here, we take the wave function in Eq.(\ref{pjwf2}) with only one SD included. Then, using an algorithm in Ref.\cite{Lian24}, the  states in a rotational band can be obtained, simultaneously. The calculated ground state band energies in $^{166}$Dy from VAPSM together with the corresponding experimental data are shown in Fig.\ref{Dy166_gsb}. One can see that with spin increases, the  calculated rotational energy increases faster than the experimental one. This discrepancy indicates that the adopted jj56pnb Hamiltonian might become less effective for $^{166}$Dy since it is already far away from $^{132}$Sn. However, with this Hamiltonian, the rotational characteristic of the ground state band in $^{166}$Dy is clearly reproduced. As a preliminary calculation, we take the wave functions corresponding to the calculated energies in Fig.\ref{Dy166_gsb} and decompose them using Eq.(\ref{cbn3}).

\begin{figure}
    \centering
    \includegraphics[width=1\linewidth]{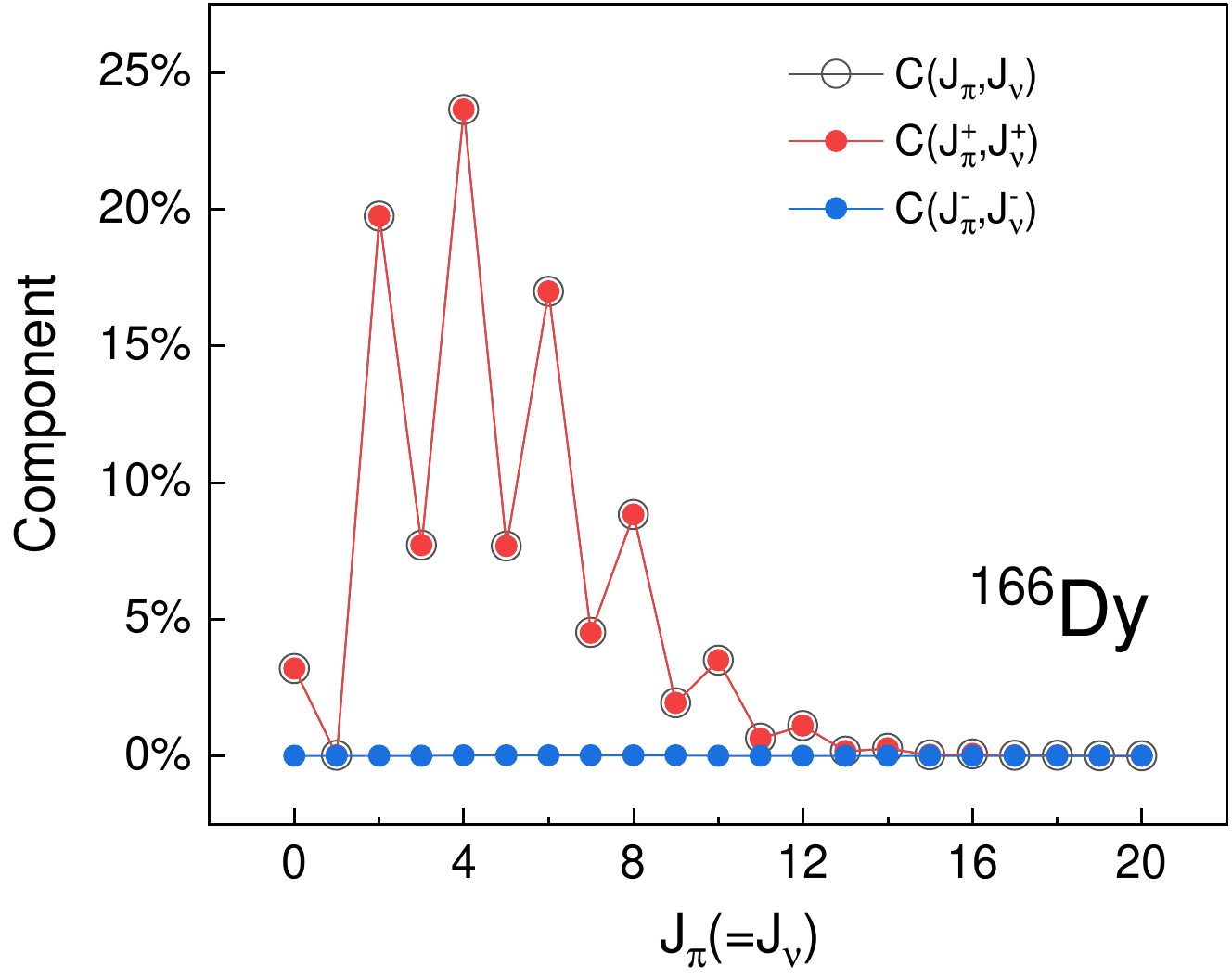}
    \caption{Distributions of the calculated $C(J_\pi,J_\nu)$, $C(J^+_\pi,J^+_\nu)$ and $C(J^-_\pi,J^-_\nu)$ values for the VAPSM wave function of the ground $0^+$ state in $^{166}$Dy.}
    \label{Dy166_0}
\end{figure}

The cutoff $J_\pi$ and $J_\nu$ values are $J_\pi=J_\nu=20$. All the calculated components with $J_\pi\leq20$ and $J_\nu\leq20$ cover 99.9998\%, 99.999\%, 99.94\% of the wave functions with $J^\pi=0^+,2^+$ and $10^+$, respectively. For the ground state with $J^\pi=0^+$, the calculated $C(J_\pi,J_\nu)$, $C(J^+_\pi,J^+_\nu)$ and $C(J^-_\pi,J^-_\nu)$ values in Eq.(\ref{cpp}) are shown in Fig.\ref{Dy166_0}. One can easily see that the $C(J^-_\pi,J^-_\nu)$ values are almost zero. Actually, the sum of all $C(J^-_\pi,J^-_\nu)$ values in Fig.\ref{Dy166_0} is about $0.042\%$. Quite different from Fig.\ref{fig1}, the predominant components in Fig.\ref{Dy166_0} are those with $J_\pi=J_\nu=2,4$ and 6, which are above 15\%, whereas $C(J_\pi=0,J_\nu=0)$ is only about 3.2\%. This result is very similar to that of $^{156}$Sm with an axial intrinsic state in Ref.\cite{otsuka1993}. But here, the adopted SD can be arbitrarily deformed. In this $J=0$ case, those nonzero $J_\pi$ and $J_\nu$ angular momenta must be anti-parallelized, as has been intensively discussed in Refs.\cite{otsuka1993,Tajima11,Tajima13}.

\begin{figure}
    \centering
    \includegraphics[width=1\linewidth]{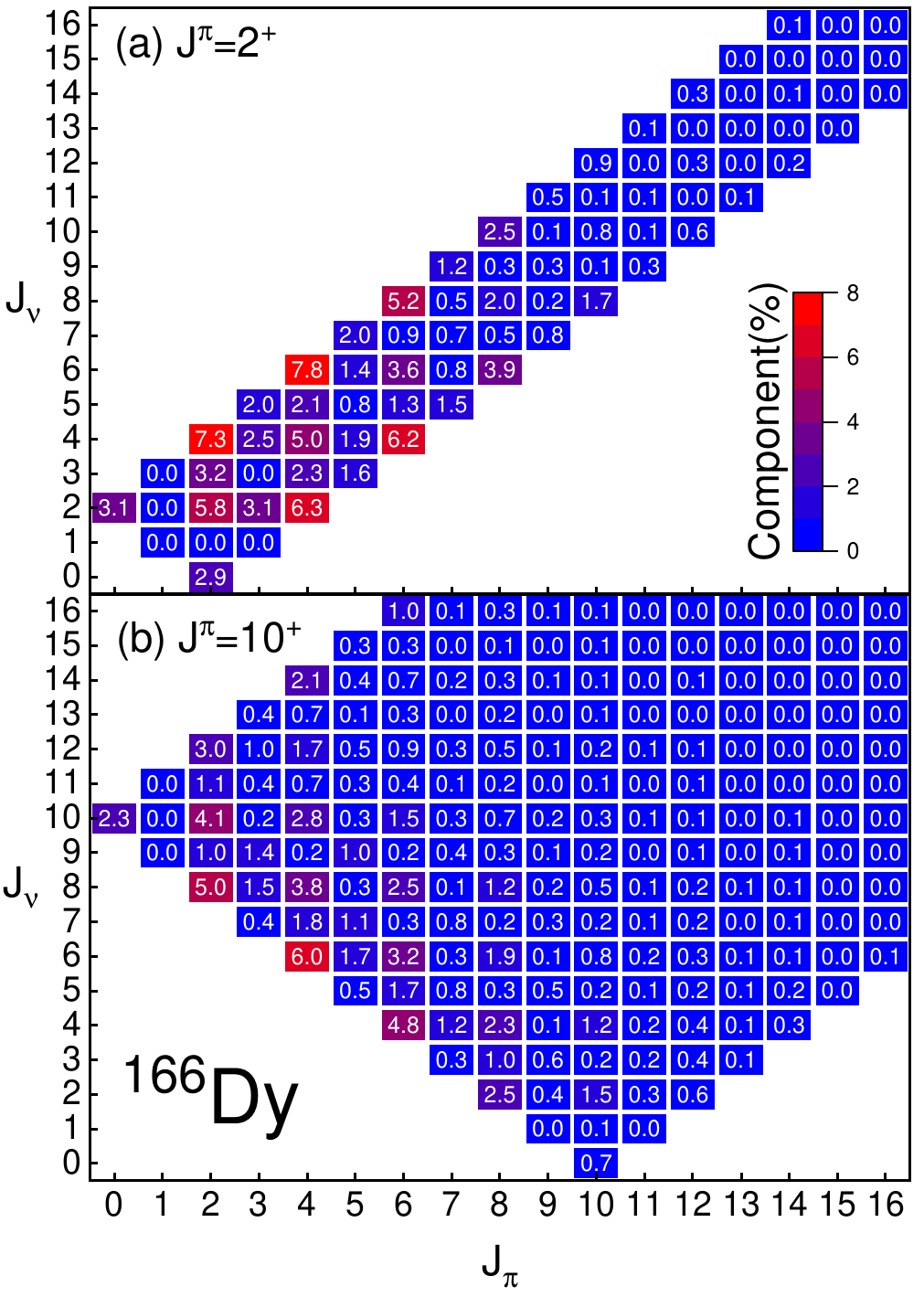}
    \caption{Distributions of the calculated $C(J_\pi,J_\nu)$ values for the VAPSM wave functions of the $2^+$ and $10^+$ states in $^{166}$Dy.}
    \label{Dy166_2}
\end{figure}

For other calculated nonzero spin states, the $C(J^-_\pi,J^-_\nu)$ values are also very close to zero. For example, at $J^\pi=2^+$ and $10^+$, the sums of calculated $C(J^-_\pi,J^-_\nu)$ values are $0.049\%$ and $0.058\%$, respectively. Thus, we have $C(J_\pi,J_\nu)\approx C(J^+_\pi,J^+_\nu)$ for the present case of $^{166}$Dy. The decompositions of the VAPSM wave functions with $J^+=2^+$ and $10^+$ are shown in Fig.\ref{Dy166_2}. For $J^+=2^+$, the predominant components are those with $J_\pi(J_\nu)=$ 4 and 6, whereas the expected $J_\pi+J_\nu=2$ components are quite small. For $J^+=10^+$, the $J_\pi+J_\nu=10$ components become predominant, which implies that the proton and neutron deformed ellipsoids rotate in almost the same direction \cite{otsuka1993,Tajima11,Tajima13}.

\section{An improvement of the VAPSM wave function}\label{imvap}

According to Eq.(\ref{dpj}), the usual projected wave function in Eq.(\ref{pjwf}) can be written in a more general form
\begin{eqnarray}\label{gpj}
|\Psi_{JM}\rangle=\sum_{J_{\pi}J_{\nu}K_{\pi}K_{\nu}i}f_{J_{\pi}J_{\nu}K_{\pi}K_{\nu}i}   P^{J_{\pi}J_{\nu}}_{JM}(K_{\pi}K_{\nu})|\Phi_i\rangle.
\end{eqnarray}
where $f_{J_{\pi}J_{\nu}K_{\pi}K_{\nu}i}=f_{K i}\langle J_{\pi}K_{\pi}J_{\nu}K_{\nu}|JK\rangle$. The $f_{K i}$ coefficients as well as $|\Phi_i\rangle$ SDs can be obtained by performing the VAPSM calculations based on Eq.(\ref{pjwf}). However, if we treat the $f_{J_{\pi}J_{\nu}K_{\pi}K_{\nu}i}$ coefficients as free parameters, namely, we recalculate $f_{J_{\pi}J_{\nu}K_{\pi}K_{\nu}i}$ by diagonalizing the Hamiltonian in the space spanned by the $  P^{J_{\pi}J_{\nu}}_{JM}(K_{\pi}K_{\nu})|\Phi_i\rangle$ bases, the shell model approximation of the  VAPSM might be further improved. For brevity, we rewrite Eq.(\ref{gpj}) in a more concise form
\begin{eqnarray}\label{gpja}
|\Psi_{JM}\rangle=\sum_{\alpha i}f_{\alpha i}P_{JM}(\alpha)|\Phi_i\rangle,
\end{eqnarray}
where $\alpha$ stands for the numbers $(J_{\pi},J_{\nu},K_{\pi},K_{\nu})$ and 
\begin{eqnarray}
 P_{JM}(\alpha)\equiv P^{J_{\pi}J_{\nu}}_{JM}(K_{\pi}K_{\nu}). 
\end{eqnarray}
Here, we adopt the simplified wave function in Eq.(\ref{pjwf1}) to perform the VAPSM calculations. Similar to Eq.(\ref{gpj}), Eq.(\ref{pjwf1}) can be written as
\begin{eqnarray}\label{gpjk}
|\Psi_{JM}(K)\rangle=\sum_{\alpha i}f_{\alpha i}   P_{JM}(\alpha )|\Phi_i\rangle,
\end{eqnarray}
with $f_{\alpha i}=f_{i}\langle J_{\pi}K_{\pi}J_{\nu}K_{\nu}|JK\rangle$. But here, $K_{\pi}$ and $K_{\nu}$ are no longer independent because they must satisfy $K_{\pi}+K_{\nu}=K$ and $K$ is fixed.

To find the fully optimized $f_{\alpha i}$ coefficients and the corresponding energy, $E$, one should solve the following generalized eigenvalue equation
\begin{eqnarray}\label{geq}
\sum_{\alpha' i'}(H_{\alpha i,\alpha' i'}-E N_{\alpha i,\alpha' i'})f_{\alpha' i'}=0
\end{eqnarray}
where $\alpha'$ stands for the numbers  $(J'_{\pi},J'_{\nu},K'_{\pi},K'_{\nu})$ and
\begin{eqnarray}
&&N_{\alpha i,\alpha' i'}=\langle \Phi_i|P^\dagger_{JM}(\alpha)P_{JM}(\alpha')|\Phi_{i'}\rangle\nonumber\\
&=&\sum_{M_\pi,M_\nu,M'_\pi,M'_\nu}\langle J_\pi M_\pi J_\nu M_\nu|JM\rangle\langle J_\pi'M_\pi'J_\nu'M_\nu'|JM\rangle\nonumber\\
    &&\times\langle\Phi^\pi_i|{P}_{K_\pi M_\pi}^{J_\pi}{P}_{M_\pi'K_\pi'}^{J_\pi'}|\Phi^\pi_{i'}\rangle\langle\Phi^\nu_i|{P}_{K_\nu M_\nu}^{J_\nu}{P}_{M_\nu'K_\nu'}^{J_\nu'}|\Phi^\nu_{i'}\rangle\nonumber\\
    &=&\langle\Phi^\pi_i|{P}_{K_\pi K_\pi'}^{J_\pi}|\Phi^\pi_{i'}\rangle\langle\Phi^\nu_i|{P}_{K_\nu K_\nu'}^{J_\nu}|\Phi^\nu_{i'}\rangle\delta_{J_\pi J_\pi'}\delta_{J_\nu J_\nu'},
\end{eqnarray}
To calculate the $H_{\alpha i,\alpha' i'}$ matrix element, the shell model Hamiltonian needs to be divided into three parts
\begin{eqnarray}\label{Hnp}
 H= H^\pi+ H^\nu+ H^{\pi\nu}.
\end{eqnarray}
 In Eq.(\ref{Hnp}), the part for like nucleons, $H^\tau$ ($\tau=\pi$ or $\nu$), can be written as 
\begin{eqnarray}\label{Hpi}
&& H^\tau=\sum_{r_\tau}\epsilon_{r_\tau} \hat N_{r_\tau}+\sum_{{r_\tau}\leq{s_\tau},{t_\tau}\leq{u_\tau},J}[\\
&&\frac{V_{{r_\tau}{s_\tau}{t_\tau}{u_\tau}}^J}{\sqrt{(1+\delta_{r_\tau s_\tau})(1+\delta_{t_\tau u_\tau})}} \sum_M  A^{\dagger}_{JM}(r_\tau s_\tau) A_{JM}(t_\tau u_\tau)].\nonumber
\end{eqnarray}
Here $\hat N_{r_\tau}$ is the operator measuring the number of nucleons in the ${r_\tau}$ subshell with quantum numbers ($n_{r_\tau}$,$l_{r_\tau}$,$j_{r_\tau}$), i.e.,
\begin{eqnarray}
\hat N_{r_\tau}=\sum_{m_{r_\tau}}  c^\dagger_{r_\tau m_{r_\tau}} c_{r_\tau m_{r_\tau}}. 
\end{eqnarray}
where $ c^\dagger_{r_\tau m_{r_\tau}}$ ($ c_{r_\tau m_{r_\tau}}$) is the spherical harmonic oscillator single particle creation (annihilation) operator with quantum numbers ($n_{r_\tau}$,$l_{r_\tau}$,$j_{r_\tau}$,$m_{r_\tau}$). 
$ A^{\dagger}_{JM}(r_\tau s_\tau)$ is the nucleon pair creation operator defined  as 
\begin{eqnarray}
 A^{\dagger}_{JM}(r_\tau s_\tau)=\sum_{m_{r_\tau}m_{s_\tau}}\langle j_{r_\tau}m_{r_\tau}j_{s_\tau}m_{s_\tau}|JM\rangle  c^\dagger_{r_\tau m_{r_\tau}} c^\dagger_{s_\tau m_{s_\tau}}\nonumber\\
\end{eqnarray}
and 
\begin{eqnarray}
 A_{JM}(t_\tau u_\tau)=\sum_{m_{t_\tau}m_{u_\tau}}\langle j_{t_\tau}m_{t_\tau}j_{u_\tau}m_{u_\tau}|JM\rangle  c_{u_\tau m_{u_\tau}} c_{t_\tau m_{t_\tau}}.\nonumber\\
\end{eqnarray}
The part for neutron-proton interaction, $ H^{\pi\nu} $, can be written as
\begin{eqnarray}
     H^{\pi\nu}=\sum_{r_{\pi}s_{\nu}t_{\pi}u_{\nu},J}V_{r_{\pi}s_{\nu}t_{\pi}u_{\nu}}^{J}\sum_{M}A_{JM}^{\dagger}(r_{\pi}s_{\nu})A_{JM}(t_{\pi}u_{\nu}),\nonumber\\
\end{eqnarray}
where
\begin{eqnarray}
 A^{\dagger}_{JM}(r_\pi s_\nu)=\sum_{m_{r_\pi}m_{s_\nu}}\langle j_{r_\pi}m_{r_\pi}j_{s_\nu}m_{s_\nu}|JM\rangle  c^\dagger_{r_\pi m_{r_\pi}} c^\dagger_{s_\nu m_{s_\nu}}\nonumber\\
\end{eqnarray}
and 
\begin{eqnarray}
 A_{JM}(t_\pi u_\nu)=\sum_{m_{t_\pi}m_{u_\nu}}\langle j_{t_\pi}m_{t_\pi}j_{u_\nu}m_{u_\nu}|JM\rangle  c_{u_\nu m_{u_\nu}} c_{t_\pi m_{t_\pi}}.\nonumber\\
\end{eqnarray}
Accordingly, the matrix element $H_{\alpha i,\alpha' i'}$ can be separated into three parts
\begin{eqnarray}
H_{\alpha i,\alpha' i'}&=&\langle \Phi_i|P^\dagger_{JM}(\alpha) H P_{JM}(\alpha')|\Phi_{i'}\rangle\nonumber\\
&=&H^\pi_{\alpha i,\alpha' i'}+H^\nu{}_{\alpha i,\alpha' i'}+H^{\pi\nu}_{\alpha i,\alpha' i'},
\end{eqnarray}
where
\begin{eqnarray}\label{hmpi}
&&H^\pi_{\alpha i,\alpha' i'}=\langle \Phi_i|P^\dagger_{JM}(\alpha) H^\pi P_{JM}(\alpha')|\Phi_{i'}\rangle\\
&=&\langle\Phi^\pi_i|{H}^\pi{P}_{K_\pi K_\pi'}^{J_\pi}|\Phi^\pi_{i'}\rangle\langle\Phi^\nu_i|{P}_{K_\nu K_\nu'}^{J_\nu}|\Phi^\nu_{i'}\rangle\delta_{J_\pi J_\pi'}\delta_{J_\nu J_\nu'},\nonumber
\end{eqnarray}
\begin{eqnarray}\label{hmnu}
&&H^\nu_{\alpha i,\alpha' i'}=\langle \Phi_i|P^\dagger_{JM}(\alpha) H^\nu P_{JM}(\alpha')|\Phi_{i'}\rangle\\
&=&\langle\Phi^\pi_i|{P}_{K_\pi K_\pi'}^{J_\pi}|\Phi^\pi_{i'}\rangle\langle\Phi^\nu_i|{H}^\nu{P}_{K_\nu K_\nu'}^{J_\nu}|\Phi^\nu_{i'}\rangle\delta_{J_\pi J_\pi'}\delta_{J_\nu J_\nu'}\nonumber
\end{eqnarray}
and
\begin{eqnarray}\label{mepn}
&&H^{\pi\nu}_{\alpha i,\alpha' i'}=\langle \Phi_i|P^\dagger_{JM}(\alpha) H^{\pi\nu} P_{JM}(\alpha')|\Phi_{i'}\rangle.
\end{eqnarray}
To evaluated $H^{\pi\nu}_{\alpha i,\alpha' i'}$,  $H^{\pi\nu}$ needs to be transformed from the particle-particle form into the particle-hole form. Namely,
\begin{equation}\label{hpinu}
\begin{split}
    H^{\pi\nu}=&\sum_{r_{\pi}s_{\nu}t_{\pi}u_{\nu},J}V_{r_{\pi}s_{\nu}t_{\pi}u_{\nu}}^{J}\sum_{M}A_{JM}^{\dagger}(r_{\pi}s_{\nu})A_{JM}(t_{\pi}u_{\nu}),\\
        =&\sum_{r_{\pi}s_{\nu}t_{\pi}u_{\nu},\lambda}\omega_{r_{\pi}t_{\pi}s_{\nu}u_{\nu}}^{\lambda}\sum_{\mu}(-1)^{\lambda-\mu}S_{\lambda\mu}(r_{\pi}t_{\pi})S_{\lambda-\mu}(s_{\nu}u_{\nu}),\\
        =&\sum_{r_{\pi}s_{\nu}t_{\pi}u_{\nu},\lambda}\omega_{r_{\pi}t_{\pi}s_{\nu}u_{\nu}}^{\lambda}(-1)^{\lambda}S_{\lambda}(r_{\pi}t_{\pi})\cdot S_{\lambda}(s_{\nu}u_{\nu}),\\
\end{split}
\end{equation}
where
\begin{equation}
\begin{split}
    \omega_{r_{\pi}t_{\pi}s_{\nu}u_{\nu}}^{\lambda}=&\sum_{J}(-1)^{j_{s_{\nu}}+j_{t_{\pi}}-\lambda-J}(2J+1)\\
    &\times\begin{Bmatrix}
    j_{r_{\pi}} & j_{s_{\nu}} & J\\ j_{u_{\nu}} & j_{t_{\pi}} & \lambda
    \end{Bmatrix}V_{r_{\pi}s_{\nu}t_{\pi}u_{\nu}}^{J},
\end{split}
\end{equation}
\begin{equation}
\begin{split}
    V_{r_{\pi}s_{\nu}t_{\pi}u_{\nu}}^{J}=&\sum_{\lambda}(-1)^{j_{s_{\nu}}+j_{t_{\pi}}-\lambda-J}(2\lambda+1)\\
    &\times \begin{Bmatrix}
    j_{r_{\pi}} & j_{s_{\nu}} & J\\ j_{u_{\nu}} & j_{t_{\pi}} & \lambda
    \end{Bmatrix}\omega_{r_{\pi}t_{\pi}s_{\nu}u_{\nu}}^{\lambda}.
\end{split}
\end{equation}
$S_{\lambda\mu}(r_\tau t_\tau)$ is a spherical tensor operator which is defined as
\begin{equation}
\begin{split}
    S_{\lambda\mu}(r_\tau t_\tau )=&\sum_{m_{r_\tau}m_{t_\tau}}\langle j_{r_\tau}m_{r_\tau}j_{t_\tau}m_{t_\tau}|\lambda\mu\rangle 
     c^\dagger_{r_\tau m_{r_\tau}}{\tilde{c}}_{t_\tau m_{t_\tau}},
\end{split}
\end{equation}
where ${\tilde{c}}_{t_\tau m_{t_\tau}}=(-1)^{j_{t_\tau}+m_{t_\tau}}{c}_{t_\tau, -m_{t_\tau}}$. Inserting Eq.(\ref{hpinu}) into Eq.(\ref{mepn}), we have
\begin{equation}\label{hnpm}
\begin{split}
    &H^{\pi\nu}_{\alpha i,\alpha' i'}\\
    =&\sum_{r_{\pi}s_{\nu}t_{\pi}u_{\nu},\lambda}\omega_{r_{\pi}t_{\pi}s_{\nu}u_{\nu}}^{\lambda}(-1)^{\lambda}\\
    &\times\langle \Phi_i|P^\dagger_{JM}(\alpha)S_{\lambda}(r_{\pi}t_{\pi})\cdot S_{\lambda}(s_{\nu}u_{\nu}) P_{JM}(\alpha')|\Phi_{i'}\rangle\\
    =&\sum_{r_{\pi}s_{\nu}t_{\pi}u_{\nu},\lambda}\omega_{r_{\pi}t_{\pi}s_{\nu}u_{\nu}}^{\lambda}(-1)^{\lambda}(-1)^{J_{\pi}'+J_{\nu}+J}\\
    &\times\begin{Bmatrix}
        J_{\pi} & J_{\nu} & J\\ J_{\nu}' & J_{\pi}' &\lambda
    \end{Bmatrix}\langle \psi_{J_{\pi}}(K_{\pi} i)||S_{\lambda}(r_{\pi}t_{\pi})||\psi_{{J'_\pi}}(K'_{\pi} i')\rangle\\
    &\times\langle \psi_{J_{\nu}}(K_{\nu}i)||S_{\lambda}(s_{\nu}u_{\nu})||\psi_{{J'_\nu}}(K'_{\nu}i')\rangle,
\end{split}
\end{equation}
where, we define
\begin{eqnarray}
| \psi_{J_\tau M_\tau}(K_\tau i)\rangle=P^{J_\tau}_{M_\tau K_\tau}|\Phi^\tau_i\rangle
\end{eqnarray}
and the reduced matrix element
\begin{eqnarray}
\langle jm|T_{\lambda\mu}|j'm'\rangle&=&\frac1{\sqrt{2j+1}}(-1)^{2\lambda}\nonumber\\
&&\times \langle j'm'\lambda\mu|jm\rangle\langle j||T_\lambda||j'\rangle.
\end{eqnarray}
Then the reduced matrix element in Eq.(\ref{hnpm}) can be expressed as
\begin{equation}\label{hmpn}
\begin{split}
    &\langle \psi_{J_{\tau}}(K_{\tau} i)||S_{\lambda}(r_{\tau}t_{\tau})||\psi_{{J'_\tau}}(K'_{\tau} i')\rangle\\
    =&\sqrt{2J_\tau+1}\sum_{\mu}\langle J'_\tau K_\tau-\mu\lambda\mu|J_\tau K_\tau\rangle\\
    &\times\langle\Phi^\tau_i|S_{\lambda\mu}(r_\tau t_\tau){P}_{K_\tau-\mu,K'_\tau}^{J'_\tau}|\Phi^\tau_{i'}\rangle.
\end{split}
\end{equation}
The calculations of the projected matrix elements in Eqs.(\ref{hmpi}), (\ref{hmnu}) and (\ref{hmpn}) can be easily established based on the present VAPSM framework. This enables us to calculate the matrix elements in Eq.(\ref{geq}), so that such generalized eigenvalue equation can be solved.

\begin{figure}
    \centering
    \includegraphics[width=1\linewidth]{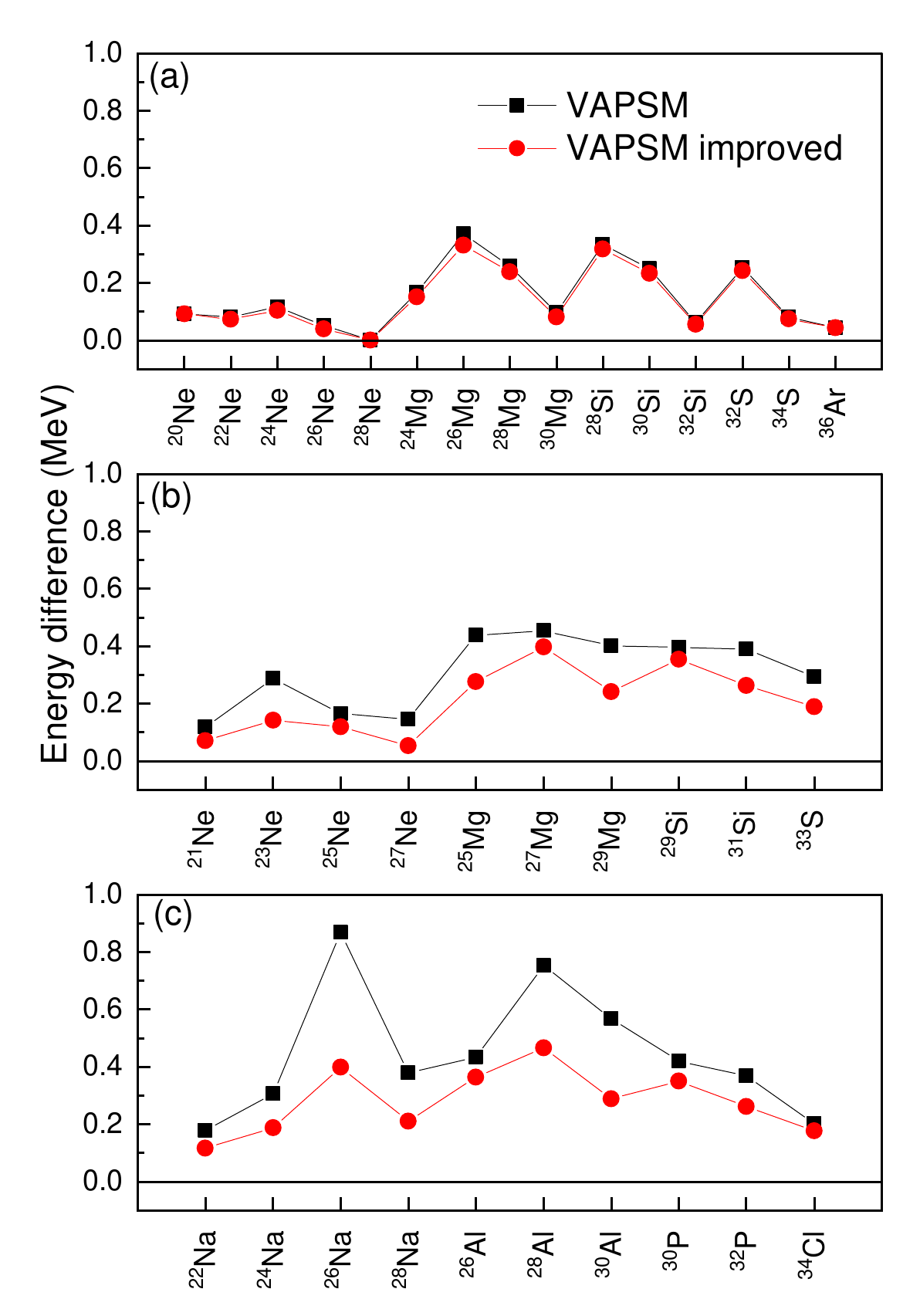}
    \caption{Calculated energy differences between the VAPSM energies and the exact SM ones for the ground states in (a) even-even, (b) odd-mass and (c) odd-odd $sd$ shell nuclei. The black squares show the results corresponding to the optimized VAPSM wave function of Eq.(\ref{pj}). The red dots show the results calculated by changing the optimized Eq.(\ref{pj}) to the generalized form of Eq.(\ref{gpjk}). The USDB interaction is adopted.}
    \label{fig:placeholder}
\end{figure}

To show the improvement of the new projected wave function. We first perform the VAPSM calculations for the ground states in some $sd$ shell nuclei. For simplicity, only a single SD is adopted to construct the VAPSM wave function in Eq.(\ref{pjwf1}). After the VAPSM calculation, one can get the minimized energy and the optimized VAPSM wave function.  The energy differences between such VAPSM energies and the exact shell model ones are shown in Fig.\ref{fig:placeholder}. One can see that these energy differences are about a few hundred keV, which seem large in this figure but actually can be negligible when looking at the absolute energies\cite{Tuya17,Wangjiaqi2018}. The good shell model approximation can also be seen from Figs.\ref{fig1},\ref{mg24},\ref{mg25},\ref{al26} and \ref{al26_0}, which indicate that the VAPSM wave functions with one SD are very close the exact shell model ones. Of course, one can reduce such energy differences to sufficiently small values by adding more and more SDs to the VAPSM wave functions. But here, we do not add more SDs to the VAPSM wave function. Instead, we are curious if such energy differences can also be reduced by adopting the aforementioned coupled projected bases. 
As a preliminary calculation, we simply use the same SD as that in the optimized VAPSM wave function to construct Eq.(\ref{gpjk}). The coefficients, $f_{\alpha i}$, and the corresponding energy can be obtained by solving Eq.(\ref{geq}). The energy differences between the  energies with new wave functions and the exact shell model ones are also shown in Fig.\ref{fig:placeholder}. One can see that for the ground states in even-even nuclei, the improvements with new wave functions are not significant. This indicates that the relative scissors motion between the neutron system and the proton system can be neglected in this calculated states. However, for odd-mass and odd-odd nuclei, the energies with new wave functions are considerably lowered. This indicates that the coupled projected basis is indeed capable of improving the VAPSM wave function, but such improvement is not complete because the adopted SD is simply taken from the original VAPSM calculation. A full VAPSM calculation with the couple projected basis is directly varying the wave function in Eq.(\ref{gpjk}), i.e., changing the $|\Phi_i\rangle$ SDs and the $f_{\alpha i}$ ) coefficients, simultaneously, which will be done in the near future.


\section{Summary and Outlook} \label{summary}

Angular momentum projection is a fundamental technique in constructing high quality nuclear wave functions which can be very close to the exact solution of a given nuclear shell model Hamiltonian.
So far, the angular momentum projected nuclear wave function is usually expressed in terms of the bases generated by performing the angular momentum projection on the reference states for the whole nuclear system. This implies the assumption that there is no collective movement of the neutrons relative to the protons.

However, the discovery of the scissors mode in nuclei confirms the extra degree of freedom for the relative movement between neutrons and protons. To treat the scissors mode, the PSM method introduced a different kind of projected bases, which are constructed by coupling the angular momentum projected wave functions for neutrons with those for the protons. Due to the  inclusion of the extra degree of freedom for scissors mode, we expect that the new projected bases may also be useful in improving the shell model approximation in our VAPSM calculations.

In this work, we first derive a general identity, which provides an explicit expansion of the angular momentum projection operator for the whole nuclear system in terms of the products of the angular momentum projection operators for the neutrons and protons. Consequently, we are enabled to expand a traditional angular momentum projected nuclear wave function in terms of the coupled projected bases, as used in the studies of the scissors mode \cite{Sun1998,Lv2022}. This is crucial for us to extract the possibility of a given $(J_\pi, J_\nu)$ pair in a given angular momentum projected nuclear wave function, i.e., the $C(J_\pi, J_\nu)$ value, as has been implemented by the standard shell model. In other words, we are able to decompose any angular momentum projected nuclear wave function into different  $(J_\pi, J_\nu)$ components.

To show the validity of such decomposition, we take the fully optimized angular momentum projected wave functions obtained from the VAPSM and decompose them into different $(J_\pi, J_\nu)$ components. Calculations for some $sd$ shell nuclei are performed. It is shown that the $C(J_\pi, J_\nu)$ distributions for the VAPSM wave functions are generally very close to those for the corresponding shell model ones. This again confirms the good shell model approximation of the VAPSM. It is shown that the distribution of $C(J_\pi, J_\nu)$ is also helpful in understanding the structure of the nuclear states. We expect that when the VAPSM is applied to a heavy deformed mass region where the standard shell model cannot reach, such decomposition might also be useful for studying the structure of heavy deformed nuclei.

Finally, we improve the VAPSM wave function by first replacing the original basis states with the coupled projected ones, then diagonalizing the shell model Hamiltonian on the new bases. For even-even nuclei, the improvements seem not so obvious, while for the odd-mass and odd-odd nuclei, the improvements are somewhat significant. This encourages us to implement a full optimization for the wave function in Eq.(\ref{gpjk}) in the future. We hope such wave function could be more close to that from the shell model and hope it could be used to study the scissors mode in heavy deformed nuclei based on a realistic shell model Hamiltonian.

\section*{ACKNOWLEDGMENTS}
This work is supported by the funding
of China Institute of Atomic Energy under Grant
No.010280825702 and the Continuous-Support Basic Scientific Research Project. We acknowledge support by the computer server SCATP in China Institute of Atomic Energy.

\bibliography{apssamp}

\end{document}